\documentstyle[aps,multicol,epsf]{revtex}

\begin{document}
\draft
\title{Anomalous Behavior of the Contact Process with Aging}

\author{
S.N. Dorogovtsev
$^{1, 2,\ast}$
and 
J.F.F. Mendes
$^{1,\dag}$
}
\address{
$^{1}$ Departamento de F\'\i sica and Centro de F\'\i sica do Porto, Faculdade de Ci\^encias, 
Universidade do Porto\\
Rua do Campo Alegre 687, 4169-007 Porto, Portugal\\
$^{2}$ A.F. Ioffe Physico-Technical Institute, 194021 St. Petersburg, Russia 
}

\maketitle

\begin{abstract}
The effect of power-law aging on a contact process is studied by simulation and using 
a mean-field approach. We find that the system may approach its stationary state in a 
nontrivial, nonmonotonous way. For the particular value of the aging 
exponent, $\alpha=1$, we observe a rich set of behaviors: depending on the process parameters, 
the relaxation to the stationary state proceeds as $1/\ln t$ or via a power law with a
nonuniversal exponent. Simulation results suggest that for 
$0<\alpha<1$, the absorbing-state phase transition is in the universality class 
of directed percolation.   
\end{abstract}

\pacs{
PACS numbers: 
64.60.Lx, 02.50.-r, 05.50.+q, 05.70.Ln
}

\begin{multicols}{2}

\narrowtext


The study of systems with absorbing-state phase transitions has been
a topic of intense research in recent years 
\cite{liggett,mdbook,h00}. 
Among these systems, the contact process (CP), introduced by T.E. Harris \cite{h74} as a model 
for spreading of disease, has become the prototype model \cite{munoz5,jd93,gt79,alon,d96}.  
In fact, it is a dynamical version of directed percolation \cite{h00,bl,essam,kinzel,j81,grass_conj,g84,cs80}. 
Recently, much effort has been directed toward generalizing the original models of 
absorbing-state phase transitions by introducing many absorbing states,
and toward finding paths to self-organized 
critical phenomena \cite{dmvz99,mdhm94,mgd98}. Therefore, one can hardly expect that an 
ordinary CP will reveal new striking feartures. 
Nevertheless, in the present Letter, we demonstrate that even the simple CP with a single 
absorbing state shows anomalous behavior if one introduces aging into the model. 
One should note that effects of aging, in the particular  
case of spreading of diceases were considered by Bernoulli \cite{ber} long time ago.
One may find a great number of models of evolution dynamics with aging in the book 
by Hoppensteadt \cite{hop}.

We show that the system with power-law aging, before it approaches its stationary state, 
behaves nonmonotonously: 
the time derivative of the particle density may change sign during the evolution. 
The case of the aging exponent, $\alpha$, equal to one is especially interesting. 
In this marginal situation, in a mean-field approach, we find different types of 
behavior. 
Depending on the parameters of the system, the density of particles changes 
at long times according to a power law with nonuniversal exponents, or proportional to $1/\ln t$. 
Such slow relaxation is certainly not typical of absorbing-state phase transitions with 
a unique absorbing state. At their critical point, relaxation normally follows a power 
law \cite{mdbook,h00}.     

We simulate the aging contact process (ACP) using sequential dynamics. Each site of the 1D 
lattice may be empty or filled by one particle. At each unit of time a particle is chosen at 
random. We decide to annihilate this particle (a) or create a new one 
(b) with probabilities $P_a(s) = 1/[ 1+p(s+1)^{\alpha} ]$ or 
$P_c(s) = p(s+1)^{\alpha}/[ 1+p(s+1)^{\alpha} ]$ correspondingly. Here, $s$ is the age of the 
particle and $\alpha$ is the aging exponent. In the case of annihilation, (a), the particle 
is deleted and we proceed to the next step increasing time by one
unit. In case (b), we chose one of two nearest neighbour sites with equal probability. 
If this site is filled, creation is impossible, and we increase time and proceed 
to the next step. If it is empty, we create a new particle at this
site, increase time, and pass to the next step. 
Thus, as compared with the ordinary CP, we 
introduce the age-dependent probabilities $P_a(s)$ and $P_c(s)$. For $\alpha > 0$, the 
case studied here, $P_a(s)$ tends to zero and $P_c(s)$ 
approaches 1 at long times.

The process is started with a random configuration of particles. 
Simulation times are about 
$10^8$ Monte-Carlo steps. The lattice size is taken up to $10^6$ sites, to minimize the 
fluctuations of the particle density, $n(t)$, which is the main quantity of interest.
The simulation results are presented in Figs. 1--3. First, one sees that at each 
$\alpha$ value studied, $n(t)$ may behave nonmonotonically, depending on the 
initial conditions and parameters of the process. If, for instance, the stationary state
is $n(\infty)=1$ (see Figs. 2 and 3), the density may first decay nearly to zero, stay in this 
range a long time, and only approach the $n(\infty)=1$ much later. 
Frequently, it is only with long-time simulations 
that we can observe the final stage of the relaxation. 

Second, in the case of $0<\alpha<1$, the critical characteristics are the same as for the 
contact process without aging, i.e., $\alpha=0$. In Fig. 1, we show
the stationary particle density
as a function of the deviation 
from the critical point, $p_c$, for two different values of the aging exponent, $\alpha=0.25$ 
and $0.75$. In both cases the data suggest that the critical exponent $\beta$ 
is not $\alpha$-dependent, and agrees with the directed percolation value, $\beta \simeq 0.277$.  
When $\alpha>1$, the phase transition is absent and the only possible stationary state is 
$n(\infty)=1$; the system exhibits an exponential approach to this limit (see. Fig. 2).

In the marginal case, $\alpha=1$, we observe a complicated, slow evolution with several 
temporal regimes characterized by 
different behaviors. Due to the slow relaxation, it is difficult to fix the final 
state. In Fig. 3 we present the results for the temporal
evolution of the particle density for several values of the parameter $p$. Depending on 
$p$, different types of the evolution are observed.     

Let us present the simplest possible mean-field description of the ACP. 
While the model is 
even simpler than that used in our simulations,
it is reasonable to expect that it can explain the principal features
observed in the latter.   
As in the ordinary CP, each site of a lattice may be filled by one particle or be empty. Each 
particle has its age, $s$. 
At each increment of time, all lattice sites are updated according to the 
following rules. 
(i) The probability for a particle of age $s$ to die at the next instant is $p_s$,
irrespective of its environment. 
(ii) The probability for a particle to survive until the next update is 
therefore $1-p_s$. 
(iii) The probability that a particle will be born at an empty site in the next instant is 
$(1/n_S)\sum_{i=1}^{n_F} q_{s(i)}$, where $n_S$ is the total number of nearest-neighbor sites, and 
$n_F$ is the total number of particles at such sites. 
$s(i)$ is the age of the particle at site $i$. 
(iv) The probability for an empty site to remain empty until the next update is therefore
$1-(1/n_S)\sum_{i=1}^{n_F} q_{s(i)}$. 
In this way we present a natural generalization of the CP,
introducing age-dependent death and birth probabilities, $p_s$ and $q_s$, respectively.

Let us derive the mean-field equations, introducing
the following quantities: the total number of particles at time $t$, 
$n_t= \sum_{s=0}^t a_{t,s}$, where $a_{t,s}$ is the number of particles of the age $s$ at time 
$t$. 
The initial condition is 
$a_{\,0,0} = n(0)$, 
where $n(0)$ is the initial number of particles.

Since the only possibility for an old particle ($s>0$) to survive is process (ii), 
the first equation is 
$a_{t+1,s+1} = (1-p_s)a_{t,s} $, 
i.e., a fraction $(1-p_s)$ of the particles with age $s$ survive. 
One may verify the solution  
$a_{t,s} =  a_{t-s,0}\prod_{u=0}^{s-1} (1-p_u)$ ,  
where $\prod_{u=0}^{-1} \equiv 1$ by definition.  

We demonstrate the derivation 
of the second equation in the simplest case of only two nearest neigbouring sites
(one may show that other coordination numbers lead to the same equation).  Particles are 
only created via process (iii). Hence,

\begin{eqnarray}
\label{4}
a_{t+1,0} & = & (1-n_t) \sum_{s,u=0}^t a_{t,s} a_{t,u} \frac{q_s + q_u}{2} +
\nonumber 
\\
& & 2 (1-n_t)^2 \sum_{s=0}^t a_{t,s} \frac{q_s}{2}   \, .
\end{eqnarray}
Thus,

\begin{equation}
\label{5}
a_{t+1,0} =  (1-n_t) \sum_{s=0}^t a_{t,s}  q_s \, .
\end{equation}  
The previous equations describe completely the ACP in the mean-field 
approach.

Applying $\sum_{s=0}^{t}$ to our first mean-field equation
one obtains  
$n_{t+1} - n_{t} = a_{t+1,0} - \sum_{s=0}^t p_s a_{t,s}$. 
Taking Eq. (\ref{5}) into account, we immediately obtain the equation generalizing the 
usual mean-field equation for the CP without aging:

\begin{equation}
\label{8}
n_{t+1} - n_{t} = \sum_{s=0}^t a_{t,s} [(1-n_t)q_s - p_s ]     \, .
\end{equation} 

Now we see that all quantities may be expressed through the $a_{t,0}$, i.e., through the 
number of particles born at instant $t$. The equation for this variable follows from Eq. 
(\ref{5}): 

\begin{eqnarray}
\label{9}
a_{t+1,0} & = & \left[1 - \sum_{s=0}^t a_{t-s,0} \prod_{u=0}^{s-1}(1-p_u) \right] \times
\nonumber 
\\[3ex] 
& &   \sum_{s=0}^t  a_{t-s,0} q_s \prod_{u=0}^{s-1}(1-p_u)  \, .
\end{eqnarray} 

Formally speaking, Eq. (\ref{9}) together with the initial condition, define the solution of our 
problem. But the problem is still too difficult for analytical treatment. 
Let us pass to the continuum 
limit. Then the equations and the initial condition become

\begin{equation}
\label{10}
n(t) = \int_0^t  ds \, a(t,s) + a(t,t)   \, , \ \ \ \  a(0,0) = n(0)    \, ,
\end{equation} 

\begin{equation}
\label{12}
\left[\frac{\partial }{\partial t} + \frac{\partial }{\partial s} + p(s) \right] a(t,s) = 0    
\, ,
\end{equation} 

\begin{eqnarray}
\label{13}
\frac{\partial n(t)}{\partial t} & = & 
[1-n(t)]
\left[ \int_0^t  ds \, a(t,s) q(s) + a(t,t)q(t) \right] -
\nonumber 
\\[3ex] 
& & \int_0^t ds \, a(t,s) p(s) + \frac{\partial a(t,t)}{\partial t}    \, ,
\end{eqnarray}  

\begin{equation}
\label{14}
a(t,0) = [1-n(t)] \left[ \int_0^t ds \, a(t,s) q(s) + a(t,t)q(t) \right]   \, .
\end{equation} 
One may check that the equations describing the continuous limit are self-consistent.
Eq. (\ref{12}) is the {\em Von Foerster} equation  \cite{murrbook}, well known in 
populational dynamics.

The solution of Eq. (\ref{12}) is 

\begin{equation}
\label{15}
a(t,s) = a(t-s,0) \exp\left[-\int_0^s du\, p(u)  \right]    \, ,
\end{equation}
so
$a(t,t) = n(0) \exp\left[-\int_0^t ds \,p(s)   \right] $ .

Here we consider only the simplest case of $q=const$ and an $s$-dependent $p(s)$. From Eq. 
(\ref{14}) we obtain the expression:
$ a(t,0) = q [1-n(t)] n(t)$.
This allows us to obtain, from Eq. (\ref{10}), and accounting for Eq. (\ref{15}), 
the following closed 
integral equation for the total number of particles:

\begin{eqnarray}
\label{20}
n(t) & = & n(0) \exp \left[ -\int_0^t ds\, p(s) \right] + 
\nonumber 
\\[3ex] 
& &  q \int_0^t ds \,n(s)[1-n(s)]  \exp \left[ -\int_0^{t-s} du\, p(u) \right]
\, .
\end{eqnarray}
Differentiating Eq. (\ref{20}) one obtains

\begin{eqnarray}
\label{200}
& &  \frac{\partial n(t)}{\partial t} = [q-p(t)]n(t) - q n^2 (t) + 
\nonumber 
\\[3ex] 
& &  q\! \int\limits_0^t \! ds \,n(s)[1-n(s)][p(t)-p(t-s)]  
\exp\! \left[ -\! \int\limits_0^{t-s} \! du\, p(u) \right]
\, . 
\nonumber
\\
& &
\end{eqnarray}
If we set $p = const$, Eq. (\ref{200}) reduces to the usual equation \cite{mdbook,h00} 
for the CP without aging. 
We write out the known results on linear relaxation for it for later comparison: 
if $p>q$, $n(\infty)=0$ and $n(t) \propto \exp[-(p-q)t]$; 
for $p<q$, $n(\infty)= 1 - p/q$ and $n(t) - n(\infty) \propto \exp[-(q-p)t]$; and 
at $p=q$, $n(t) \propto 1/t$. 
 
The reasonable dependences for $p(s)$ and $q(s)$, which admit comparison with our simulations, 
are the following: $p(s)$ decreases gradually as the particle age $s$ increases, and $q(s)$ 
increases with $s$ increasing or is constant. 
Here, we consider the annihilation probability $p(s) = c (s+t_0)^{-\alpha}$, where $\alpha$ is 
the aging exponent. 
The constant $c$ plays the same role as the parameter $p^{-1}$ in our simulations. 
One can check that, in the particular case of $q=0$, i.e., when creation of particles is absent, 
Eq. (\ref{20}) is valid for any dimension, and exact. 
In this case, at $0<\alpha<1$, 
$n(t)$ approaches zero, 
$n(t)=n(0)\exp[t_0^{1-\alpha}/(1-\alpha)]\exp[-(t+t_0)^{1-\alpha}/(1-\alpha)]$,
at $\alpha>1$, 
$n(t)$ quickly approaches a constant value 
$n(\infty) = n(0) \exp\{-c t_0^{\alpha-1}/(\alpha-1) \}$,
and, for $\alpha=1$, $n(t)=n(0)t_0/(t+t_0)$.

Eq. (\ref{20}) may be easily solved by iteration. One may start, e.g., from $n^{(i)}(t) = 
n(0)$. 
The resulting solutions $n(t)$ are very similar to those obtained by numerical simulation; we do 
not present these curves here, but only describe the results of the direct analysis of Eq. 
(\ref{20}).  

First of all, expanding Eq. (\ref{20}) for small $t$ one finds that

\begin{equation}
\label{20a}
\frac{\partial n(0)}{\partial t} = n(0) \left\{q[1-n(0)] - c t_0^{-\alpha}  \right\}   \, .
\end{equation}
(One may check that this relation is also valid for an $s$-dependent $q(s)$ if one replaces $q$ 
by $q(0)$.)
Hence, $n(t)$ may increase or decrease from $n(0)$ at short times depending on a particular 
relation between the constants of the problem and on the initial condition. In principle, the 
sign of the derivative is independent of the value $n(\infty)$. That 
leads to the observed nonmonotonic behavior of $n(t)$. 

The nonmonotonous behaviour observerd in some regimes can be intuitively understood regarding that in the very begining 
the particles are by necessity young,
so most of them easily die off, but later the survivors
turn to be nearly immortal, and the lattice will be filled up with old-timers.

Let us now consider the behavior of $n(t)$ for different values of $\alpha$ at long times. 
(Note that all the following asymptotes may be obtained from the Laplace transform of Eq. 
(\ref{20}) linearized near the corresponding stationary state.) 

{\it Regime I}, $0<\alpha<1$. The situation is very similar to the CP without aging. 
There is a critical point $c^\ast(\alpha,q,t_0)$ for each value of $\alpha$ and $q$.  
At $c>c^\ast(\alpha,q,t_0)$, $n(\infty)=0$ and $n(t)$ has an exponentional approach
to the stationary state. 
At the critical point, $c=c^\ast(\alpha,q,t_0)$, $n(t)$ relaxes to zero by a power law, $n(t) 
\propto 1/t$.
At $c<c^\ast(\alpha,q,t_0)$, at long times, $n(t)$ approaches $n(\infty)>0$ exponentially, 
$n(t)-n(\infty) \propto \exp\{-g(\alpha,c,q,t_0) t\}$. Here, $g(\alpha,c,q,t_0)$ is zero at the 
critical point, and also approaches zero when $\alpha$ approaches $1$.

{\it Regime II}, $\alpha>1$. For any $c$ and $q$, $n(t)$ approaches $1$ exponentially. 
The critical point (and scaling) are absent. 
The reasons are the following. At $\alpha>1$, the kernel $\exp\{-\int_0^t du c(u+t_0)^{-\alpha} 
\}$ quickly decreases to a constant value, 
$\exp\{-c t_0^{\alpha-1}/(\alpha-1) \}$, as $t$ grows, and one can substitute this constant into 
Eq. (\ref{20}). 
Taking the derivative of Eq. (\ref{20}) and linearizing the resulting equation near 
$n(\infty)=1$ we get

\begin{equation}
\label{21}
1-n(t) \propto \exp \left\{       
-q \exp[-c t_0^{-(\alpha-1)}/(\alpha-1)]\, t
\right\}   
\, .
\end{equation}

{\it Regime III}, $\alpha=1$, the most intriguing case. Several types of critical behavior are 
realized for different values of $c$.

(a) For $c<1$ and any $q$, $n(\infty)=1$ and the long-time dependence is $1-n(t) \propto 
t^{-(1-c)}$. 
Depending on the relation between $c, q$, and $t_0$ [see Eq. (\ref{20a})], $n(t)$ exhibits this 
dependence immediately or, on the contrary, 
$n(t)$ first decreases, remains near zero for some time, and then very slowly approaches $1$.
The kernel $\exp\{-\int_0^t ds\,p(s) \}=[(t+t_0)/t_0]^{-c}$, so if we demand $n(t \to \infty) 
\to 1$ in Eq. (\ref{20}), we immediately obtain such a form for the asymptote. 

(b) For $c=1$ and any $q$, $n(\infty)=1$ and the long-time behavior is $1-n(t) \propto 1/\ln t$. 

(c) For $1<c<1+q t_0$, $n(\infty) = 1 - (c-1)/(q t_0)$ and 
$n(t)-n(\infty) \propto t^{-c}$. 

(d) At $c = 1+q t_0$, $n(\infty)=0$ and $n(t) \propto 1/t$. 
The kernel is a quickly decreasing function as compared with $n(t)$, 
so the reason for such behavior is the same as for the $1/t$ relaxation 
at the critical point at $0 \leq \alpha < 1$.

(e) At $c \geq 1+q t_0$ (note that the definition of $p(s)$ imposes the additional restriction 
on possible values of $c$ 
and $t_0$: $c<t_0$), $n(\infty)=0$ and  $n(t) \propto t^{-c} $. 

The exponents $1-c$ and $c$ are {\em nonuniversal} since $c$ is simply a coefficient in the 
definition of the annihilation probability. 
Previously,
nonuniversality was found 
in models with multiple absorbing states \cite{munoz5,jd93,d96,mdhm94,mgd98,lm97} and in the 
process of spreading in media with memory \cite{gcr97}. Nevertheless, the kind of nonuniversality 
observed here was not seen.  

One should note that the results for {\it Regime III} were obtained only in the frame of
mean-field theory. We did not study the role of fluctuations, and our simulations do not let us 
fix the final 
stationary state (except for the case $n(\infty)=1$), when the relaxation is slow, i.e., for 
$\alpha=1$. Note also that our mean-field approach is based on parallel dynamics while a 
sequentional one was used in the numerical simulations. This difference, however,
seems to be not crucial for the CP with a single absorbing state. 
 
In summary, we have studied the influence of power-law aging on the contact process with a 
single absorbing state. Both simulation and mean-field theory reveal the 
nonmonotonic character of the relaxation. In the marginal case of a unit aging 
exponent, in the frame of mean-field theory, we found that the relaxation of the particle 
density to the stationary state proceeds by a power-law with a nonuniversal exponent, or even 
slower, i.e., proportional to $1/\ln t$. 
\\

SND thanks PRAXIS XXI (Portugal) for a research grant PRAXIS XXI/BCC/16418/98. JFFM was 
partially supported by the project PRAXIS/2/2.1/FIS/299/94. We  
thank R. Dickman for interesting and stimulating discussions and for useful comments on the manuscript. 
We also thank M.C. Marques and M.A. Santos for useful discussions.
\\
$^{\ast}$      Electronic address: sdorogov@fc.up.pt\\
$^{\dagger}$   Electronic address: jfmendes@fc.up.pt

\newpage

\begin{figure}
\epsfxsize=85mm
\epsffile{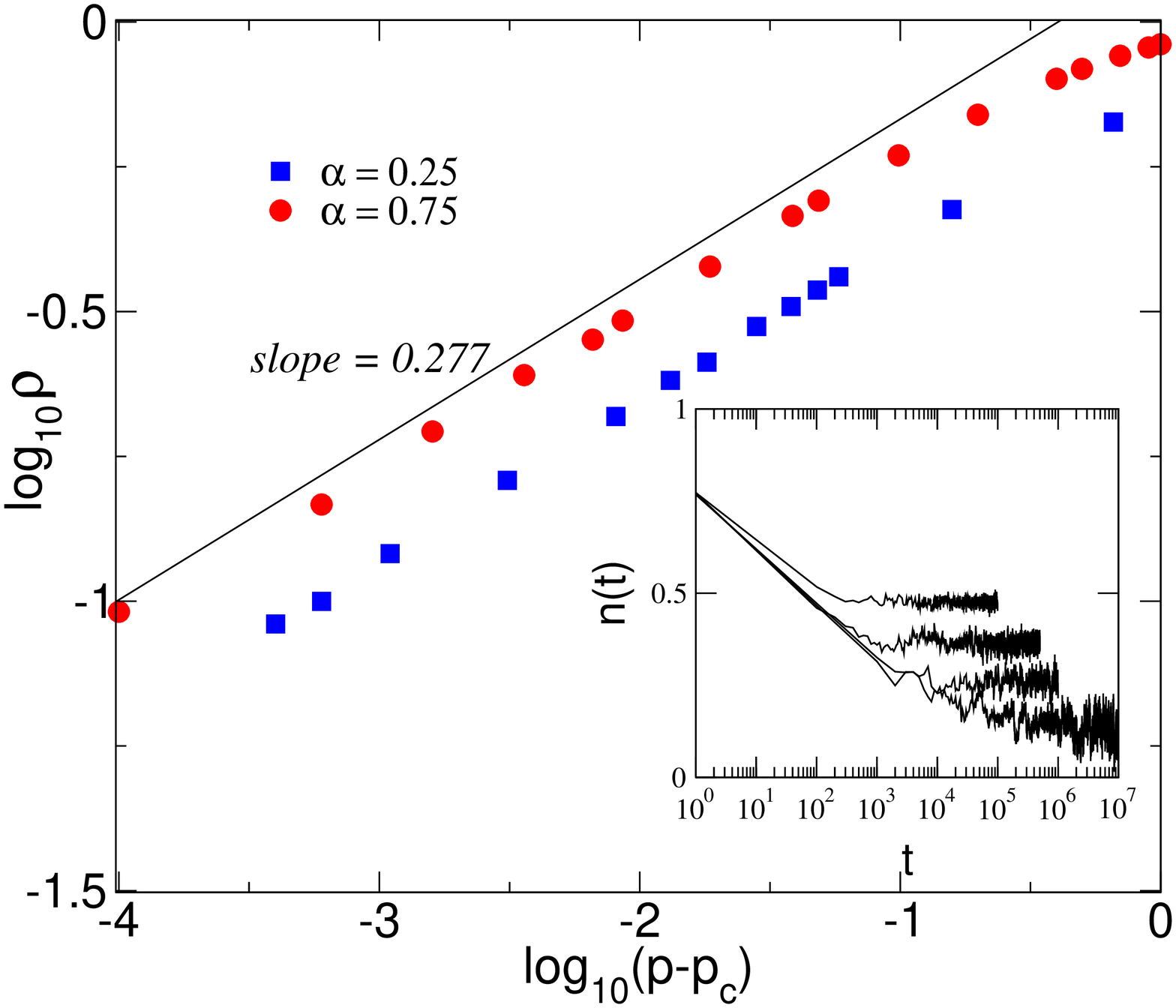}
\caption{
Log-log plot of the stationary density $\rho$ vs. $p-p_c$ for two different values of the aging 
exponent $\alpha$, $0.25$ and $0.75$. 
The critical points are $p_c(0.25)=2.3419(3)$ and $p_c(0.75)=1.1014(3)$.
The full line has a slope corresponding to the $\beta$ exponent of DP in the 1+1 dimension. The 
inset shows the evolution of $n(t)$ for different values of $p$ (from top to below, 
$p=2.5, \ 2.4, \ 2.36$, and $2.3423$) and $\alpha=0.25$.
}
\label{f1}
\end{figure}

\begin{figure}
\epsfxsize=85mm
\epsffile{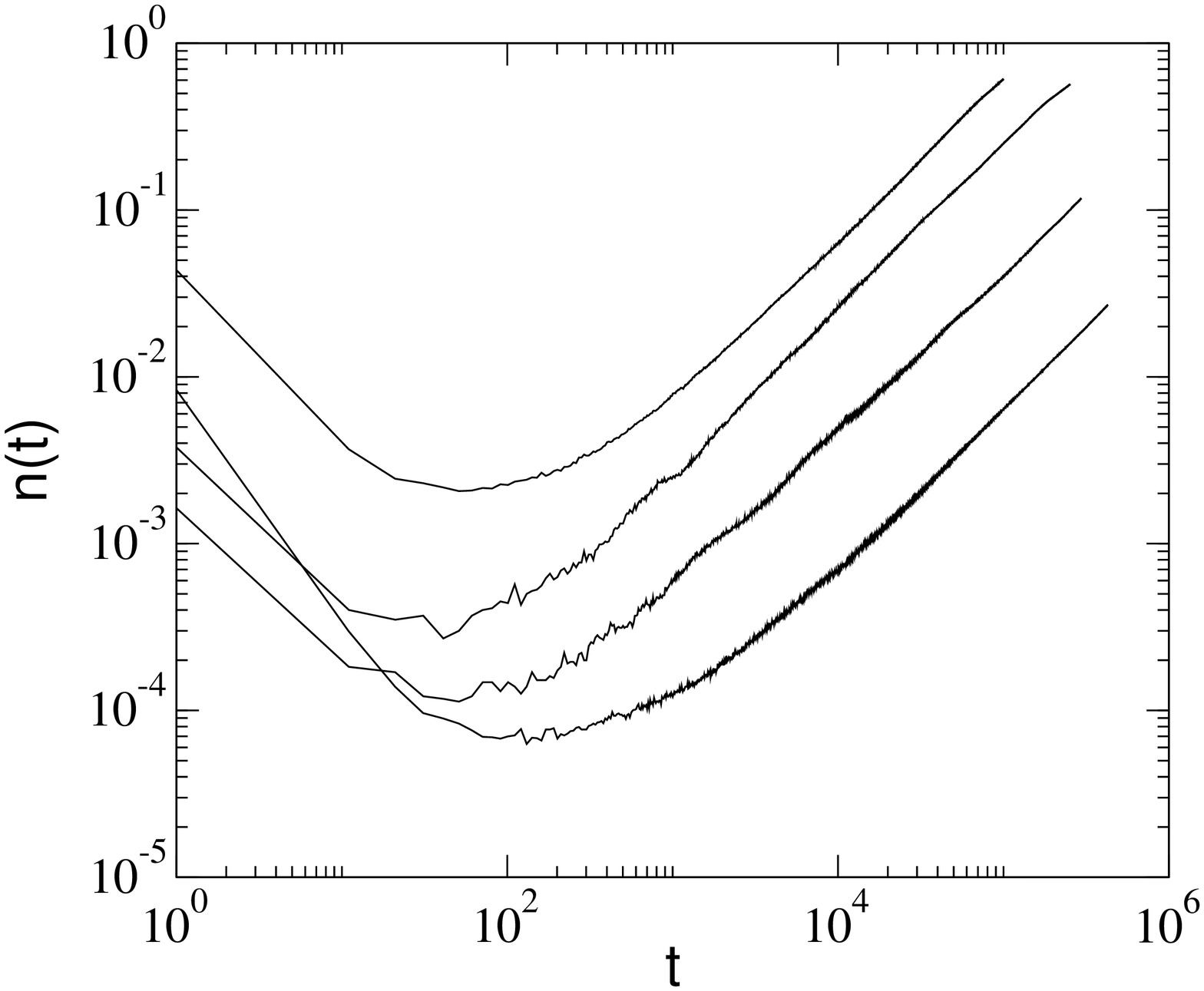}
\caption{
Evolution of $n(t)$ for different values of $p$ and initial densities, for $\alpha=2$.  
From top to below, $p=0.075, \ n_0=0.01; \ p=0.075, \ n_0=0.25;\ p=0.05, \ n_0=0.05;$ and $p=0.1, \ n_0=0.01$. 
}
\label{f2}
\end{figure}

\begin{figure}
\epsfxsize=85mm
\epsffile{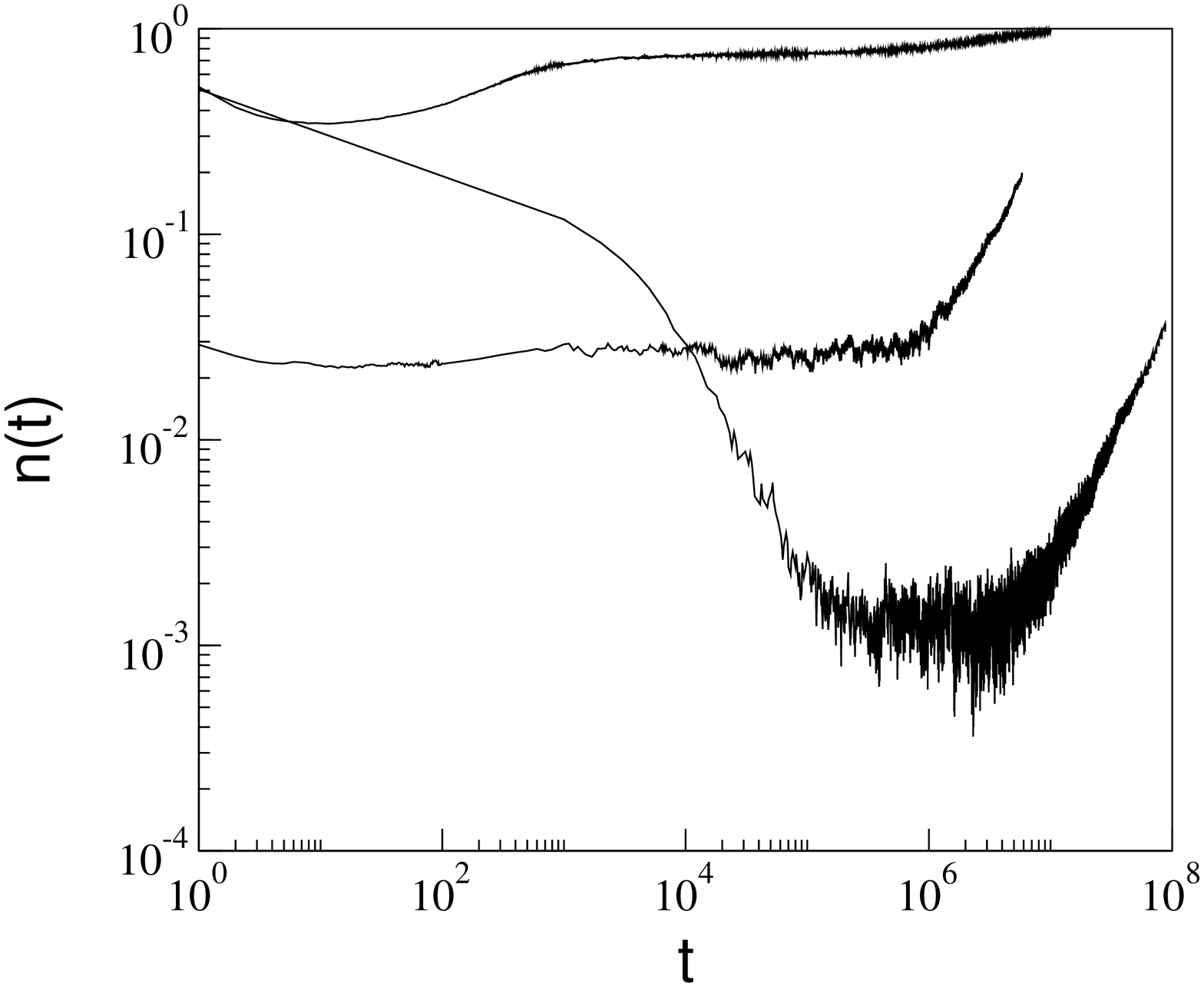}
\caption{
Temporal evolution of the particle density, $n(t)$, for three different values of $p$ and 
$\alpha=1$. From top to below, $p=0.65, \ 0.6667$, and $0.75$.
}
\label{f3}
\end{figure}

\end{multicols}


\begin{references}

\bibitem{liggett} 
T.M. Liggett, Interacting Particle Systems 
(Springer-Verlag, New York,1985).

\bibitem{mdbook} J. Marro and R. Dickman, {\it Nonequilibrium phase transitions in lattice 
models} (Cambridge University Press, Princeton, Cambridge, 1999).

\bibitem{h00}  H. Hinrichsen, cond-mat/0001070, to be published in Ann. Phys..

\bibitem{h74} T.E. Harris, Ann. Prob. {\bf 2}, 969 (1974).

\bibitem{munoz5} 
M. A. Mu\~{n}oz, G. Grinstein, R. Dickman and R. Livi, 
Phys. Rev. Lett. {\bf 76}, 451 (1996)

\bibitem{jd93} 
I. Jensen, R. Dickman, Phys. Rev. E {\bf 48}, 1710(1993).

\bibitem{gt79} 
P. Grassberger and de la Torre, Ann. Phys.  {\bf 122}, 373 (1979).

\bibitem{alon} 
M. Alon, M. R. Evans, H. Hinrichsen and D. Mukamel, Phys. Rev. Lett. {\bf 76}, 2746 (1996).

\bibitem{d96} 
R. Dickman, Phys. Rev. E {\bf 53}, 2223 (1996).

\bibitem{bl} 
J. Blease, J. Phys. C {\bf 10}, 923 (1977)

\bibitem{essam} 
J. W. Essam and K. De'Bell, J. Phys. A {\bf 14}, L459 (1981)

\bibitem{kinzel} 
W. Kinzel, Z. Physik B {\bf 58}, 229 (1985).

\bibitem{j81} 
H. K. Jenssen, Z. Phys. {\bf B 42}, 151 (1981)

\bibitem{grass_conj}  
P. Grassberger, Z. Phys. {\bf B 47}, 365 (1982).

\bibitem{g84} 
P. Grassberger, J. Phys. A {\bf 22}, L1103 (1984).

\bibitem{cs80} 
J. L.~Cardy and R. L.~Sugar, J. Phys. A {\bf 13}, L423 (1980).

\bibitem{dmvz99} R. Dickman, M.A. Mu\~{n}oz, A. Vespignani, and S. Zapperi, cond-mat/9910454, to 
be published in Braz. J. Phys.

\bibitem{mdhm94} J.F.F. Mendes, R. Dickman, M. Henkel, and M.C. Marques, J. Phys. A  {\bf 27}, 
3019 (1994).

\bibitem{mgd98} M.A. Mu\~{n}oz, G. Grinstein, and R. Dickman, J. Stat. Phys. {\bf 91}, 541 
(1998).

\bibitem{ber} D. Bernoulli, Histoire  de l'Acad. Roy. Sci. (Paris) avec M{\'e}m. des Math. et Phys., M{\'e}m., 1 (1760)

\bibitem{hop} F.C. Hopensteadt, {\it Mathematical Theories of Populations: DEmographics, Genetics and Epidemics}. 
CBMS Lectures Vol. 20 (SIAM Publications, Philadelphia, 1975).

\bibitem{murrbook} J.D. Murray, {\it Mathematical Biology} (Springer, Berlin, 1993).

\bibitem{lm97} 
C. Lopez and M.A. Mu\~{n}oz, Phys. Rev. E {\bf 56}, 4864 (1997).

\bibitem{gcr97} P. Grassberger, H. Chat\'{e}, and G. Rousseau, Phys. Rev. E {\bf 55}, 2488 
(1997).


\end{references}
\end{document}